\documentclass[amsmath,amssymb,aps,eqsecnum]{revtex4}
\usepackage{amsmath}
\usepackage{amsfonts}
\usepackage{amssymb}
\usepackage{bm,bbm}
\usepackage{graphicx}

\usepackage{multido}   
\usepackage{pst-all}

\newcommand\inp[2]{\left\langle\,#1,#2\,\right\rangle}
\newcommand\norm[1]{\|#1\|}
\newcommand\CC{\mathbbm C}

\newcommand\one{\mathbbm 1}

\newcommand\A{\mathcal A}

\newcommand\B{\mathcal B}
\renewcommand\H{\mathcal H}

\renewcommand\L{\mathcal L}
\newcommand\M{\mathcal M}

\newcommand\ph{\varphi}
\renewcommand\a{\alpha}
\newcommand\s{\sigma}

\newcommand\tr{\hbox{tr}}
\newcommand\id{\hbox{id}}

\newcommand\set[2]{\{#1\,|\,#2\}}
\newcommand\half{\textstyle{{1\over2}}}
\newcommand\som[1]{\sum_{#1=1}^m}
\newcommand\somp[1]{\sum_{#1=1}^{m'}}

\newcommand\tuple[2]{#1_1,\ldots,#1_{#2}}
\newcommand\ketbra[1]{|#1\rangle\langle#1|}
\newcommand\Ketbra[2]{|#1\rangle\langle#2|}

\newtheorem{theorem}{Theorem}
\newtheorem{lemma}[theorem]{Lemma}
\newtheorem{proposition}[theorem]{Proposition}
\newtheorem{corollary}[theorem]{Corollary}

\begin{document}
\title {Protected Subspaces in Quantum Information}

\author{Krzysztof Majgier$^{1}$,  Hans Maassen$^{2}$, 
 Karol {\.Z}yczkowski$^{1,3}$}
\affiliation {$^1$Instytut Fizyki im. Smoluchowskiego,
 Uniwersytet Jagiello{\'n}ski,
 ul. Reymonta 4, 30-059 Krak{\'o}w, Poland}
\affiliation{$^2$Department of Mathematics, University of Nijmegen,
Heyendaalseweg 135, 
6525 AJ Nijmegen, The Netherlands}
\affiliation{$^3$Centrum Fizyki Teoretycznej, Polska Akademia Nauk,
Al. Lotnik{\'o}w 32/44, 02-668 Warszawa, Poland}

 \date{September 1, 2009}

\begin{abstract}
In certain situations the state of a quantum system,
after transmission through a quantum channel,
can be perfectly restored.
This can be done by ``coding'' the state space of the system
before transmission into a ``protected'' part of a larger state space,
and by applying a proper ``decoding'' map afterwards.
By a version of the Heisenberg Principle, which we prove,
such a protected space must be ``dark'' in the sense that
no information leaks out during the transmission.
We explain the role of the Knill-Laflamme condition in relation
to protection and darkness,
and we analyze several degrees of protection,
whether related to error correction, or to state restauration after
a measurement.
Recent results on higher rank numerical ranges of operators
are used to construct examples.
In particular, dark spaces are constructed for any map of rank 2,
for a biased permutations channel and for certain separable maps
acting on multipartite systems.
Furthermore, error correction subspaces are provided  
for a class of tri-unitary noise models.
\end{abstract}


\maketitle

\medskip


\section{Introduction}
We consider a quantum channel of finite dimension
through which a quantum system in some state is sent.
The output consists of another quantum state,
and possibly some classical information.
We are interested in the question to what extent the original quantum state
can be recovered from that state and that information.
In particular, we investigate if there are subspaces of the Hilbert
space of the original system, on which the state can be perfectly
restored.

In the literature a hierarchy of such spaces, which we shall call
{\it protected subspaces} here, has been described.
The strongest protection possible is provided in the case of a
``decoherence free subspace''  \cite{DG97,ZR97,LCW98a,SL05}.
In this case the channel acts on the subspace as a isometric transformation.
All we have to do in order to recover the state, is to rotate it back.

The next strongest form of protection occurs
when the channel acts on the subspace as a random choice between isometries,
whose image spaces are mutually orthogonal.
Then by measuring along a suitable partition of the output Hilbert space,
it can be inferred from the output state which isometry has occurred,
so that it can be rotated back.
This situation is characterized by the well-known Knill--Laflamme criterion,
\cite{BDSW96a,KL97}  
and the protected subspace in this case is usually called
an {\it error correction subspace}.

The weakest form of protection is provided in yet a third situation,
which was encountered in the context of quantum trajectories
and the purification tendency of states along these paths \cite{MK06}.
In this case the deformation of the state is not caused by some given
external device, but by the experimenter himself, who is performing a
Kraus measurement \cite{Kr71}.
Also in this case the ``channel'' acts as a random isometry,
but the image spaces need not be orthogonal.
It is now the {\it measurement outcome} (not the output state),
that betrays to the experimenter which isometry has taken place.
Using this information, he is able to undo the deformation
of the component of the state that lies in the subspace considered.

It should be emphasized that the latter form of protection is far from
a general error correction procedure.
The experimenter only repairs the damage that he himself
has incurred by his measurement.

Nevertheless, the above situations seem mathematically sufficiently similar
to deserve study under a common title.

In all these three cases the experimenter learns nothing 
during the recovery operation
about the component of the state inside our subspace.
In this sense these subspaces can be considered  ``dark'',
and this darkness is essential for the protection of information.
Our main result (Theorem \ref{TheoremProtDark}) is concerned with the
equivalence between protection and darkness,
which is a consequence of Heisenberg's principle that no information
on an unknown quantum state can be obtained without disturbing it
(Corollary \ref{HeisCor}).

The question arises,
for what channels protected subspaces are to be be found.
We consider several examples in their Kraus decompositions.
In each decomposition, we look for subspaces on which the channel
acts as a multiple of an isometry, to be called a {\it homometry} here.
Obviously, every (Kraus) operator $A$ acts homometrically on a
one-dimensional space $\CC\psi$;
its image $\CC A\psi$ is another one-dimensional space,
and the shrinking factor is $\sqrt{\inp{A\psi}{A\psi}}=\norm{A\psi}$.
However, one-dimensional spaces are useless as coding spaces for quantum
states.
What we shall need, therefore,
is the recent theory of higher rank numerical ranges
\cite{CKZ06b,CKZ06}.
With the help of this we shall be able to construct several
examples.

The paper is organized as follows. 
A brief review of basic concepts 
including  channels and instruments  
is presented in section II.
We discuss Heisenberg's principle in Section III.
and prove our main Theorem, Theorem \ref{TheoremProtDark}
in Section IV.
In subsequent sections
we analyze different forms of protected subspaces
and compare their properties.
In section V we review the notion of higher rank numerical range
and quote some results on existence in the
algebraic compression problem. Some examples
of dark subspaces are presented in section VI,
while an exemplary problem of finding an error
correction code for a specific model of tri--unitary noise
acting on a $3\times K$ system is solved in section VII.


\section{Channels and Instruments}
\label{sec:Notation}
Let $\H$ be a finite-dimensional complex Hilbert space,
and let $\B(\H)$ denote the space of all linear operators
on $\H$.
We consider $\H$ as the space of pure states of some quantum system.
By a {\it quantum operation} or {\it channel} on this system
we mean a completely positive map $\Phi:\B(\H)\to\B(\H)$ mapping
the identity operator $\one=\one_\H$ to itself.
The map $\Phi$ describes the operation ``in the Heisenberg picture'',
i.e. as an action on observables.
Its description ``in the Schr\"odinger picture'',
i.e. as an action on density matrices $\rho$,
is described by its adjoint $\Phi^*$. 
The maps $\Phi$ and $\Phi^*$ are related by
   $$\forall_\rho\forall_{X\in\B(\H)}:\quad
     \tr\bigl(\Phi^*(\rho)X\bigr)=\tr\bigl(\rho\Phi(X)\bigr)\; .$$
We note that the property $\Phi(\one)=\one$, which we require for $\Phi$,
is equivalent to trace preservation by $\Phi^*$:
   $$\tr\bigl(\Phi^*(\rho)\bigr)
        =\tr\bigl(\Phi^*(\rho)\cdot\one\bigr)
        =\tr\bigl(\rho\cdot\Phi(\one)\bigr)
        =\tr(\rho\cdot\one)
        =\tr(\rho)\;.$$
By Stinespring's theorem, every channel $\Phi:\B(\H)\to\B(\H)$
can be written as
\begin{equation}
\label{EqStinespring}
\Phi(X)=V^\dagger (X\otimes\one_\M)V\;,
\end{equation}
where $V$ is an isometry $\H\to\H\otimes\M$ for some auxiliary
Hilbert space $\M$.
The minimal dimension $r$ of $\M$ admitting such a representation
is called the {\it Choi rank} \cite{Cho75a,BZ06} of $\Phi$.

Any Stinespring representation of $\Phi$
naturally leads to a wider quantum operation
\begin{equation}
\label{EqPsiWide}
\Psi:\B(\H)\otimes\B(\M)\to\B(\H):
     X\otimes Y\mapsto V^\dagger (X\otimes Y)V\;,
\end{equation}
which can be interpreted (in the Heisenberg picture) as
the result of coupling the system to some {\it ancilla}
having Hilbert space $\M$.

\begin{figure}[ht]
\bigskip\bigskip \psset{xunit=1cm,yunit=1cm,plotpoints=200}
\begin{pspicture}(-3,-0.5)(12,0.3)
\linethickness{1pt}  

\put(1.4,-0.4){\line(1,0){0.8}}
\put(1.4,0.4){\line(1,0){0.8}}
\put(1.4,-0.4){\line(0,1){0.8}}
\put(2.2,-0.4){\line(0,1){0.8}}
\put(1.65,-0.1){$\Phi$}

\put(7.4,0){\line(4,3){0.8}}
\put(7.4,0){\line(4,-3){0.8}}
\put(8.2,-0.6){\line(0,1){1.2}}
\put(7.8,-0.1){$\Psi$}

\psplot{0}{1.4}{x 600 mul sin 0.1 mul}
\put(0.4,0.3){$\B(\H)$}
\psplot{2.2}{3.6}{x 600 mul sin 0.1 mul}
\put(2.6,0.3){$\B(\H)$}

\psplot{8.2}{10.0}{x 8 sub 600 mul sin 0.1 mul 0.3 add}
\put(8.6,0.6){$\B(\H)$}

\psplot{8.2}{9.2}{x 8 sub 600 mul sin 0.1 mul -0.3 add}
\put(8.3,-0.8){$\B(\M)$}

\psplot{6.0}{7.4}{x 8 sub 600 mul sin 0.1 mul}
\put(6.4,0.3){$\B(\H)$}

\psset{linewidth=2pt}   
\psline(9.0,-0.5)(9.4,-0.1) \psline(9.0,-0.1)(9.4,-0.5)

\linethickness{2pt}
\put(4.6,0.06){\line(1,0){0.4}}
\put(4.6,-0.06){\line(1,0){0.4}}

\end{pspicture}

\caption{Stinespring's dilation of $\Phi$ seen as coupling to an ancilla $\M$
\label{FigStinespring}}

\end{figure}
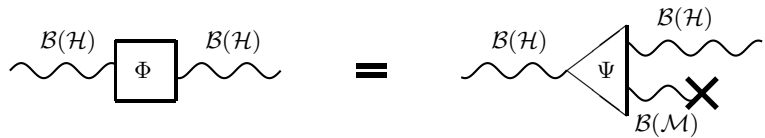

Thus Stinespring's representation (\ref{EqStinespring})
can be symbolically rendered as in Fig. \ref{FigStinespring}.


\noindent
In this picture, the cross stands for the substitution of $\one_\M$
(in the Heisenberg picture, reading from right to left),
or the partial trace
(in the Schr\"odinger picture, reading from left to right).
Physically, it corresponds to throwing away, or just ignoring,
the ancilla after the interaction.
In the picture, the fact that
$\Psi$ is a {\it compression}, i.e. $\Psi=V^\dagger\cdot V$ for some
isometry $V$, is symbolized by the triangular form of its box.

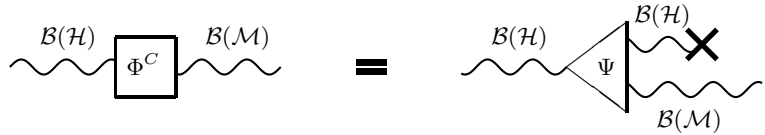
\begin{figure}[ht]
\bigskip\bigskip \psset{xunit=1cm,yunit=1cm,plotpoints=200}
\begin{pspicture}(-3,-0.5)(12,0.3)
\linethickness{1pt}  

\put(1.4,-0.4){\line(1,0){0.8}}
\put(1.4,0.4){\line(1,0){0.8}}
\put(1.4,-0.4){\line(0,1){0.8}}
\put(2.2,-0.4){\line(0,1){0.8}}
\put(1.55,-0.1){$\Phi^C$}

\put(7.4,0){\line(4,3){0.8}}
\put(7.4,0){\line(4,-3){0.8}}
\put(8.2,-0.6){\line(0,1){1.2}}
\put(7.8,-0.1){$\Psi$}

\psplot{0}{1.4}{x 600 mul sin 0.1 mul}
\put(0.4,0.3){$\B(\H)$}
\psplot{2.2}{3.6}{x 600 mul sin 0.1 mul}
\put(2.6,0.3){$\B(\M)$}

\psplot{6.0}{7.4}{x 8 sub 600 mul sin 0.1 mul}
\put(6.4,0.3){$\B(\H)$}

\psplot{8.2}{10.0}{x 8 sub 600 mul sin 0.1 mul 0.3 sub}
\put(8.3,0.6){$\B(\H)$}

\psplot{8.2}{9.2}{x 8 sub 600 mul sin 0.1 mul 0.3 add}
\put(8.6,-0.8){$\B(\M)$}

\psset{linewidth=2pt}   
\psline(9.0,0.5)(9.4,0.1) \psline(9.0,0.1)(9.4,0.5)

\linethickness{2pt}
\put(4.6,0.06){\line(1,0){0.4}}
\put(4.6,-0.06){\line(1,0){0.4}}

\end{pspicture}

\caption{The conjugate channel $\Phi^C$.
\label{FigConjugate}}

\end{figure}

\noindent
Now, by blocking the other exit in Fig. \ref{FigStinespring},
we obtain the conjugate channel \cite{KNMR},
$\Phi^C$:
   $$\Phi^C:\B(\M)\to\B(\H):\quad Y\mapsto\Psi(\one_\H\otimes Y)
                                 =V^\dagger(\one_\H\otimes Y)V\;.$$
See also Fig. \ref{FigConjugate}.


\noindent
The main message of this paper is the following.
The conjugate channel can be viewed as the flow of information
into the environment.
By Heisenberg's Principle, to be explained below,
such a flow prohibits the faithful
transmission of information through the original channel $\Phi$.
In particular, if the information encoded in some subspace of $\H$
is to be transmitted faithfully, nothing of it is visible
from the outside: protection implies darkness.
The degree of protection (decoherence free, strong or weak) 
is related to the degree of darkness,
for which we shall define some terminology.

\begin{figure}[ht]
\psset{xunit=1cm,yunit=1cm,plotpoints=200}
\begin{pspicture}(-3.0,-0.5)(7,0.3)


\linethickness{1pt}
\put(1.6,-0.7){\line(0,1){0.8}}
\linethickness{2pt}
\put(1.6,-0.7){\line(2,1){0.85}}
\put(1.65,-0.7){\line(2,1){0.85}}
\put(1.6,0.1){\line(2,-1){0.85}}
\put(1.65,0.1){\line(2,-1){0.85}}
\put(2.45,-0.7){\line(0,1){0.8}}
\put(1.7,-0.4){$\Pi_f$}
\put(0.5,0){$\B(\M)$}
\put(3.1,0){$\CC^m$}

\linethickness{1pt}
\put(2.4,-0.3){\line(1,0){1.6}}

\psplot{0}{1.6}{x 4.4 sub 600 mul sin 0.1 mul -0.3 add}

\end{pspicture}
\caption{Von Neumann measurement on $\M$.}
\label{FigNeumann}

\end{figure}
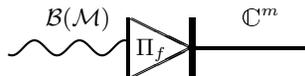

Any orthonormal basis $f=(\tuple f m)$ in $\M$
corresponds to a possible von Neumann measurement $\Pi_f^*$
on the ancilla,
which maps a density matrix $\rho$ on $\M$ to
a probability distribution $(\inp{f_1}{\rho f_1},\inp{f_2}{\rho f_2},
\ldots,\inp{f_m}{\rho f_m})$ on $\{1,2,\ldots,m\}$.
(Cf. Fig. \ref{FigNeumann}.)
In the Heisenberg picture this is the map from the algebra $\CC^m$
with generators $e_1=(1,0,0,\ldots,0)$, $e_2=(0,1,0,\ldots,0)$, $\ldots$,
$e_m=(0,0,\ldots,0,1)$, to $\B(\M)$, given by
   $$\Pi_f:\quad e_i\mapsto\ketbra{f_i}\;.$$

In FIG. \ref{FigNeumann} the abelian algebra $\CC^m$ is indicated by
a straight line since it only carries classical information.
Quantum information is designated by a wavy line.

\noindent
Let us now denote by $I_f$ the ``partial inner product map''
   $$\H\otimes\M\to\H:\quad \ph\otimes\theta\mapsto\inp f\theta\ph\;,$$
and let us write
   $$A_i:=I_{f_i}V\quad\in\B(\H)\;.$$
Then since $I_{f_i}^\dagger X I_{f_j}=X\otimes\Ketbra{f_i}{f_j}$,
we obtain a decomposition of $\Phi$ along the basis $(f_i)_{i=1}^m$
as follows: 
\begin{equation}
\Phi(X)=\Psi(X\otimes\one_\M)=\som i \Psi\bigl(X\otimes\ketbra{f_i}\bigr)
            =\som i V^\dagger I_{f_i}^\dagger X I_{f_i} V
            =\som i A_j^\dagger XA_j\;.
\label{EqDecomp}
\end{equation}
This is a {\it Kraus decomposition} of $\Phi$.
Combining the coupling to the ancilla with a von Neumann measurement
on the latter, we obtain an {\it instrument} in the language of Davies
and Lewis \cite{DaL}:
\begin{equation}
   \Psi_f:\B(\H)\otimes\CC^m\to\B(\M):\quad
      X\otimes e_i\mapsto V^\dagger(X\otimes\ketbra{f_i})V
     =A_i^\dagger X A_i\;.
\label{EqInstrument}
\end{equation}


\noindent
The isometric property of $V$ is now expressed as
\begin{equation}
\label{EqIsometry}
V^\dagger V=\som i A_i^\dagger A_i=\one\;.
\end{equation}

\section{Heisenberg's Principle or Observer Effect}
In quantum mechanics observables are represented as self-adjoint operators
on a Hilbert space.
When $A$ and $B$ are commuting operators,
then they possess a common complete orthonormal set of eigenvectors.
Each of these eigenvectors $\psi$ determines a state which associates sharply
determined values to both observables $A$ and $B$.

But when $A$ and $B$ do not commute, such states may not exist.
This important property of quantum mechanics was first discussed by
Heisenberg \cite{Hei27},
and is called the {\it Heisenberg Uncertainty Principle}.
It was formulated by Robertson \cite{Rob29} in the form
   $$\s_\psi(A)\cdot\s_\psi(B)\ge\half|\inp\psi{(AB-BA)\psi}|\;.$$
Here $\s_\psi(X)$ is the standard deviation
of $X$ in the distribution induced by $\psi$.
Already in the very same paper,
Heisenberg introduced a second and very different principle,
which is sometimes designated as the ``Observer Effect'',
and which we shall call the Heisenberg Principle here.
Roughly speaking, it says that:
\begin{eqnarray}
&&\hbox{{\it if $A$ and $B$ do not commute,}}\cr
&&\hbox{{\it a measurement of $B$ perturbs the
             probability distribution of $A$.}}
\end{eqnarray}
\label{HP1}\noindent
In the first half century of quantum mechanics,
physicists, including Heisenberg himself, were satisfied with this formulation,
and even considered it more or less identical to the Uncertainty Principle
above.

In recent years it was realized that in fact we have here two
different principles.
Good quantitative formulations have been given of the Heisenberg Principle
(for example \cite{Wer04,Jan06}).
For the purpose of the present paper we are satisfied with
a qualitative ('yes-or-no') version.

Let us first note that the formulation of the principle needs sharpening.
As it stands, the condition is not needed:
already in the trivial case that $A=B$ measurement of $B$ changes
the probability distribution of $A$.
Indeed changing the probability distribution of an observable is the
very purpose of measurement!
And also, when $A$ and $B$ commute, but are correlated,
then gaining information on
$B$ typically changes the distribution of $A$.
A characteristic property of quantum theory only arises
if we require that the outcome of the measurement of $A$ is not used
in the determination of the new probability distribution of $B$.
Even then, some states may go through unchanged.

Corrected for these observations, the Heisenberg Principle reads:
\begin{eqnarray}
  &&\hbox{{\it For noncommuting $A$ and $B$ we cannot avoid that,}}\cr
  &&\hbox{{\it for some initial states,
a measurement of $B$ changes the distribution of $A$,}}\cr
  &&\hbox{{\it even if we ignore the outcome of the measurement.}}
\end{eqnarray}
\label{HP2}\noindent
The contraposition of the statement turns out to be mathematically more
tractible:
\begin{eqnarray}
  &&\hbox{{\it If the probability distribution of $A$ is {\rm not} altered
in any initial state}}\cr
  &&\hbox{{\it --- by us performing some measurement
and ignoring its outcome ---}}\cr
  &&\hbox{{\it then the object measured must commute with $A$.}}
\label{HP3}
\end{eqnarray}
\noindent
In this form it is sometimes called the 'nondemolition principle'.

Now let us make this statement precise.
We start with a self-adjoint operator $A$ on $\H$.
Its distribution in the state $\rho$ is determined by the numbers
$\tr(\rho g(A))$ when $g$ runs through the functions on the spectrum of $A$.
Then some quantum operation is performed which on $\B(\H)$
is described by a completely positive unit preserving map $\Phi$.
We require that for all states $\rho$ and all functions $f$
   $$\tr\bigl(\Phi^*(\rho)g(A)\bigr)=\tr\bigl(\rho g(A)\bigr)\;,$$
which is equivalent to
   $$\Phi\bigl(g(A)\bigr)=g(A)\;.$$
I.e.: all elements of the *-algebra $\A$ consisting of functions of $A$
are left invariant by $\Phi$.
Let us denote the {\it commutant}
of $\A$ by $\A'$,
\begin{equation}
  \A' = \set{X\in\B(\H)}{\forall_{Y\in\A}\;:XY=YX} \ . 
\label{commut}
\end{equation}

Now, the quantum operation $\Phi$ is due to a measurement,
so it is actually of the form
   $$\Phi(X)=\Theta(X\otimes\one),$$
where $\Theta:\B(\H)\otimes\CC^m\to\B(\H)$ is some instrument whose outcomes,
labeled $1,2,\ldots,m$, in the state $\rho$
have probabilities $p_1,p_2,\ldots,p_m$ to occur, where
   $$p_j=\tr\bigl(\rho\Theta(\one\otimes e_j)\bigr)\;,$$
and where $\tr\bigl(\rho\Theta(X\otimes e_j)\bigr)/p_j$ is the expectation
of $X$, conditioned on the outcome $j$.
(This situation is comparable to, but more general than,
that of $\Psi_f$ in (\ref{EqInstrument}).)
Here $\CC^m$ is the algebra of measurement outcomes.
Generalizing to arbitrary $\A$,
we may now formulate the Heisenberg Principle as follows.

\begin{proposition}
\label{HeisPrin}
{\bf(Heisenberg Principle.)}
Let $\H$ be a finite dimensional Hilbert space,
and $\B$ some finite dimensional *-algebra.
Let $\A$ be a sub-*-algebra of $\B(\H)$,
and let $\Theta$ be a completely positive unit preserving map
$\B(\H)\otimes\B\to\B(\H)$.
Suppose that for all $A\in\A$ we have
   $$\Theta(A\otimes\one)=A\;.$$
Then for all $B\in\B$
   $$\Theta(\one\otimes B)\in\A'\;.$$

\end{proposition}

\smallskip\noindent{\bf Proof:}
For any density matrix $\rho$ on $\H$,
define the quadratic form $D_\rho$ on $\B(\H)\otimes\B$ by
   $$D_\rho(X,Y):=\tr\rho\bigl(\Theta(X^*Y)-\Theta(X)^*\Theta(Y)\bigr)\;.$$
By the Cauchy-Schwartz inequality for the completely positive map
$\Theta$ this quadratic form is positive semidefinite.
By assumption we have for all $A\in\A$:
\begin{eqnarray*}
D_\rho(A\otimes\one,A\otimes\one)&=&
      \tr\rho\bigl(\Theta(A^*A\otimes\one)
     -\Theta(A\otimes\one)^*\Theta(A\otimes\one)\bigr)\\
     &=&\tr\rho\bigl(A^*A\otimes\one-(A\otimes\one)^*(A\otimes\one)\bigr)
      =0\;.
\end{eqnarray*}
It then follows from the Cauchy-Schwartz inequality for $D_\rho$
itself that $D_\rho(A\otimes\one,\one\otimes B)=0$ for all $B\in\B$.
But then
\begin{eqnarray*}
\tr\rho\bigl(A\Theta(\one\otimes B)\bigr)
&=&\tr\rho\bigl(\Theta(A\otimes\one)\Theta(\one\otimes B)\bigr)
 =\tr\rho\bigl(\Theta\bigl((A\otimes\one)(\one\otimes B)\bigr)\bigr)
 =\tr\rho\bigl(\Theta\bigl((\one\otimes B)(A\otimes\one)\bigr)\bigr)\\
&=&\tr\rho\bigl(\Theta(\one\otimes B)\Theta(A\otimes\one)\bigr)
 =\tr\rho\bigl(\Theta(\one\otimes B)A\bigr).
\end{eqnarray*} 
Since this holds for all $\rho$,
it follows that $\Theta(\one\otimes B)$ commutes with $A$.
{\hfill $\square$}

\smallskip\noindent
By taking $\A$ and $\B$ abelian, say $\A$ generated by some observable
$A$, and $\B=\CC^m$ as above, and by choosing for $\Theta$ some
instrument giving information about $B$, we obtain a statement
of the type (\ref{HP3}).

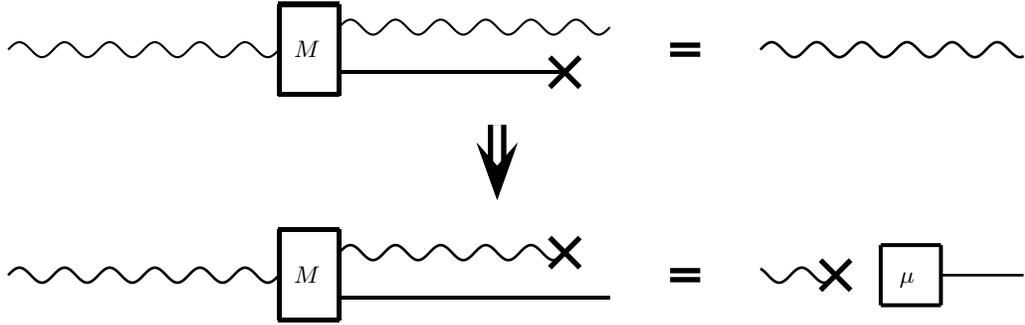
\begin{figure}
\psset{xunit=1cm,yunit=1cm,plotpoints=200}
\begin{pspicture}(0,-3.7)(12,0.3)
\linethickness{1.5pt} \put(3.6,-0.6){\line(1,0){0.8}}
\put(3.6,-0.6){\line(0,1){1.2}} \put(4.4,-0.6){\line(0,1){1.2}}
\put(3.6,0.6){\line(1,0){0.8}} \put(3.8,-0.1){$M$} \psplot{0}{3.6}{x
600 mul sin 0.1 mul} \psplot{4.4}{8}{x 8 sub 600 mul sin 0.1 mul 0.3
add} \linethickness{1pt} \put(4.4,-0.3){\line(1,0){3}}
\psset{linewidth=2pt} \psline(8.8,-0.06)(9.2,-0.06)
\psline(8.8,0.06)(9.2,0.06)
\psset{linewidth=1pt} \psplot{10}{13.5}{x 10 sub 600 mul sin 0.1
mul} \psset{linewidth=2pt} \psline(7.2,-0.5)(7.6,-0.1)
\psline(7.2,-0.1)(7.6,-0.5)
\psline[doubleline=true,linewidth=2pt,doublesep=3pt]{->}%
    (6.5,-1)(6.5,-2)
\linethickness{1.5pt} \put(3.6,-3.6){\line(1,0){0.8}}
\put(3.6,-3.6){\line(0,1){1.2}} \put(4.4,-3.6){\line(0,1){1.2}}
\put(3.6,-2.4){\line(1,0){0.8}} \put(3.8,-3.1){$M$}
\psset{linewidth=1pt} \psplot{0}{3.6}{x 600 mul sin 0.1 mul 3 sub}
\psplot{4.4}{7.4}{x 8 sub 600 mul sin 0.1 mul 2.7 sub}
\linethickness{1pt} \put(4.4,-3.3){\line(1,0){3.6}}
\psset{linewidth=2pt} \psline(8.8,-3.06)(9.2,-3.06)
\psline(8.8,-2.94)(9.2,-2.94)
\psset{linewidth=1pt} \psplot{10}{11}{x 11 sub 600 mul sin 0.1 mul 3
sub}
\psset{linewidth=2pt} \psline(7.2,-2.9)(7.6,-2.5)
\psline(7.2,-2.5)(7.6,-2.9)
\psline(10.8,-2.8)(11.2,-3.2) \psline(10.8,-3.2)(11.2,-2.8)
\psset{linewidth=1.5pt} \psline(11.6,-3.4)(11.6,-2.6)
\psline(11.6,-2.6)(12.4,-2.6) \psline(12.4,-2.6)(12.4,-3.4)
\psline(12.4,-3.4)(11.6,-3.4) \put(11.85,-3.1){$\mu$}
\psset{linewidth=1pt} \psline(12.4,-3)(13.5,-3)
\end{pspicture}
\caption{Heisenberg's Principle as an implication between diagrams}
\label{HPpicture}
\end{figure}

But there are other possible conclusions.
We may choose $\A=\B(\H)$, so that $\A'=\CC\cdot\one_\H$.
Then the Heisenberg principle says that,
if we wish to make sure that any possible state
$\rho$ on $\H$ be unchanged by our measurement,
no information at all concerning $\rho$ can be gained.
This is expressed by the following corollary and FIG. \ref{HPpicture}.

\begin{corollary}
\label{HeisCor}
In the situation of Proposition \ref{HeisPrin},
if for all $A\in\B(\H)$ we have
   $$\Theta(A\otimes\one)=A\;,$$
then there is a positive normalized linear form
$\alpha$ on $\B$ such that for all $B\in\B$:
   $$\Theta(\one\otimes B)=\alpha(B)\cdot\one_\H\;.$$
\end{corollary}

\noindent
Indeed, the expectation of an outcome observable,
   $$\tr(\Theta^*\rho)(\one\otimes B)
    =\tr(\rho\Theta(\one\otimes B))
    =\tr(\rho\one_\H)\cdot\tr(\a B)=\tr(\a B)$$
does not depend on $\rho$ (see FIG. \ref{HPpicture}.)

\bigskip


\section{Protection and Darkness: the Knill-Laflamme Condition}
Let $\L$ be a complex Hilbert space of dimension smaller than that of $\H$,
and let $C:\L\to\H$ be some isometry.
The range of $C$ is a subspace of $\H$, isomorphic with $\L$.
Let $\Gamma:\B(\H)\to\B(\L)$ denote the {\it compression map}
   $$\Gamma(X)=C^\dagger XC\;.$$
Note that $\Gamma$ is completely positive and identity-preserving.
Compression maps are a convenient way of describing subspaces of
a Hilbert space in the language of operations.
Note that the operation $\Gamma^* $ (in the Sch\"odinger picture)
embeds density matrices on $\L$ into the range of $C$:
   $$\Gamma^* (\rho)=C\rho C^\dagger\;.$$
Physically, $\Gamma$ is to be viewed as the ``coding'' operation.

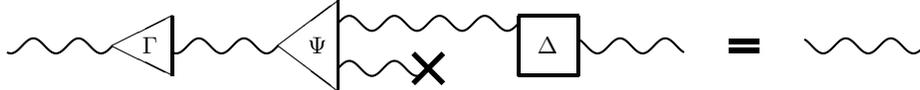
\begin{figure}[ht]
\bigskip\bigskip \psset{xunit=1cm,yunit=1cm,plotpoints=200}
\begin{pspicture}(-3,-0.5)(15,0.3)
\linethickness{1pt}  

\put(1.4,0){\line(2,1){0.8}}
\put(1.4,0){\line(2,-1){0.8}}
\put(2.2,-0.4){\line(0,1){0.8}}
\put(1.8,-0.1){$\Gamma$}

\put(3.6,0){\line(4,3){0.8}}
\put(3.6,0){\line(4,-3){0.8}}
\put(4.4,-0.6){\line(0,1){1.2}}
\put(4,-0.1){$\Psi$}

\put(6.8,-0.4){\line(1,0){0.8}}
\put(6.8,0.4){\line(1,0){0.8}}
\put(6.8,-0.4){\line(0,1){0.8}}
\put(7.6,-0.4){\line(0,1){0.8}}
\put(7.05,-0.1){$\Delta$}

\psplot{0}{1.4}{x 140 sub 600 mul sin 0.1 mul}
\psplot{2.2}{3.6}{x 600 mul sin 0.1 mul}

\psplot{4.4}{6.8}{x 8 sub 600 mul sin 0.1 mul 0.3 add}

\psplot{4.4}{5.6}{x 8 sub 600 mul sin 0.1 mul -0.3 add}

\psplot{7.6}{9.0}{x 8 sub 600 mul sin 0.1 mul}

\psplot{10.6}{12.2}{x 8 sub 600 mul sin 0.1 mul}

\psset{linewidth=2pt}   
\psline(5.4,-0.5)(5.8,-0.1) \psline(5.4,-0.1)(5.8,-0.5)

\linethickness{2pt}
\put(9.6,0.06){\line(1,0){0.4}}
\put(9.6,-0.06){\line(1,0){0.4}}

\end{pspicture}

\caption{Strong protection of $\Gamma$ against $\Psi$
\label{FigStrongProt}}

\end{figure}

\smallskip\noindent{\bf Definition.}
We say that $\Gamma$ (or the subspace $C\L$ of $\H$)
is {\it protected against} a channel $\Phi:\B(\H)\to\B(\H)$
if $\Gamma\circ\Phi$ is right-invertible,
i.e. if there exists a ``decoding'' operation
$\Delta:\B(\L)\to\B(\H)$ such that
\begin{equation}
   \Gamma\circ\Phi\circ\Delta=\id_{\B(\L)}\;.
\label{EqProtected}
\end{equation}
By virtue of (\ref{EqStinespring}) we may picture this state of affairs
as in Fig. \ref{FigStrongProt}.

\begin{figure}[ht]
\psset{xunit=1cm,yunit=1cm,plotpoints=200}
\begin{pspicture}(-3,-0.5)(15,0.3)
\linethickness{1pt}  

\put(1.4,0){\line(2,1){0.8}}
\put(1.4,0){\line(2,-1){0.8}}
\put(2.2,-0.4){\line(0,1){0.8}}
\put(1.8,-0.1){$\Gamma$}

\put(3.6,0){\line(4,3){0.8}}
\put(3.6,0){\line(4,-3){0.8}}
\put(4.4,-0.6){\line(0,1){1.2}}
\put(4,-0.1){$\Psi$}

\put(6.8,-0.6){\line(1,0){0.8}}
\put(6.8,0.6){\line(1,0){0.8}}
\put(6.8,-0.6){\line(0,1){1.2}}
\put(7.6,-0.6){\line(0,1){1.2}}
\put(7.0,-0.1){$\Delta_f$}

\psplot{0}{1.4}{x 140 sub 600 mul sin 0.1 mul}
\psplot{2.2}{3.6}{x 600 mul sin 0.1 mul}

\psplot{4.4}{6.8}{x 8 sub 600 mul sin 0.1 mul 0.3 add}

\psplot{4.4}{5.2}{x 8 sub 600 mul sin 0.1 mul -0.3 add}

\psplot{7.6}{9.0}{x 8 sub 600 mul sin 0.1 mul}

\psplot{10.6}{12.2}{x 8 sub 600 mul sin 0.1 mul}



\linethickness{2pt}
\put(9.6,0.06){\line(1,0){0.4}}
\put(9.6,-0.06){\line(1,0){0.4}}


\linethickness{1pt}
\put(5.2,-0.7){\line(2,1){0.85}}
\put(5.2,-0.7){\line(0,1){0.8}}
\put(5.2,0.1){\line(2,-1){0.85}}
\put(6.0,-0.3){\line(1,0){0.8}}
\linethickness{2pt}
\put(6.0,-0.7){\line(0,1){0.8}}
\psplot{4.4}{5.2}{x 4.4 sub 600 mul sin 0.1 mul -0.3 add}
\put(5.25,-0.4){$\Pi_f$}


\end{pspicture}

\caption{Weak protection of $\Gamma$ against $\Psi_f$
\label{FigWeakProt}}

\end{figure}
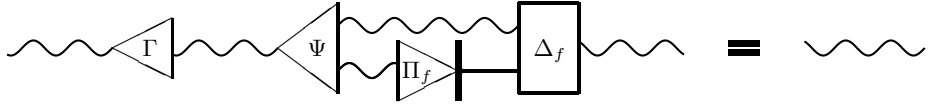

\noindent
The subspace will be called {\it weakly protected against}
an instrument $\Psi_f:\B(\H)\otimes\CC^m\to\B(\H)$
if $\Gamma\circ\Psi_f$ is right-invertible,
i.e. if there exists a decoding operation
$\Delta_f:\B(\L)\to\B(\H)\otimes\CC^m$ such that
\begin{equation}
   \Gamma\circ\Psi_f\circ\Delta_f=\id_{\B(\L)}\;.
\label{EqWeakProt}
\end{equation}
This is symbolically rendered in Fig. \ref{FigWeakProt}.
The difference with Fig. \ref{FigStrongProt} is that,
in the case of weak protection,
it is allowed to use the measurement outcome in the decoding.
In the figure the classical information
consisting of the measurement outcome,
is symbolized by a straight line.

The above notions concern protection of information.
Now we consider its availability to the external world.

\smallskip\noindent{\bf Definition.}
Let $\Psi_f:\B(\H)\otimes\CC^m\to\B(\H)$ denote a quantum measurement 
(instrument) as described in (\ref{EqInstrument}).
The subspace $C\L\subset\H$
(or the compression operation $\Gamma=C^\dagger \cdot C$),
will be called {\it dark} with respect to $\Psi_f$
if for all $i=1,\ldots,m$ we have
\begin{equation}
\Gamma\circ\Psi_f(\one\otimes e_i)\in\CC\cdot\one_\L\; .
\label{EqDark}
\end{equation}
This condition can be written in an equivalent form, 
\begin{equation}
   C^\dagger A_i^\dagger A_iC=\lambda_i\cdot\one_\L
{\rm \quad for \quad}
i=1,\dots, m\; .
\label{EqDarkKraus}
\end{equation}
\noindent
The subspace $C\L$ will be called {\it completely dark} for a channel
$\Phi:\B(\H)\to\B(\H)$ if it is dark for all Kraus measurements
$\Psi_f$ obtained by choosing different orthonormal bases in the ancilla
space of some Stinespring dilation of $\Phi$; i.e.
\begin{equation}
\forall_{Y\in\B(\M)}:
     \quad\Gamma\circ\Psi(\one\otimes Y)\in\CC\cdot\one_\L\;.
\label{EqComplDark}
\end{equation}
In terms of Kraus operators this is equivalent with
the {\sl Knill-Laflamme condition}: 
\begin{equation}
   C^\dagger A_i^\dagger A_jC=\alpha_{i,j}\cdot\one_\L
{\rm \quad for \quad}
i,j=1,\dots, m\; .
\label{EqKnillLaflamme}
\end{equation}

\medskip\noindent
{\bf Interpretation:}
From (\ref{EqDark}) and (\ref{EqDarkKraus}) we see that,
if the von Neumann measurement along $f$ is performed,
the measurement outcome $i$ 
has the same probability
$\rho\bigl(\Gamma\circ\Psi_f(\one\otimes e_i)\bigr)
=\rho(C^\dagger A_i^\dagger A_i C)=\lambda_i$,
in all system states $\rho$, 
i.e. no information concerning the state $\rho$
can be read off from the $f$-measurement on the ancilla.

\noindent
Complete darkness (i.e. (\ref{EqComplDark}) or the equivalent Knill-Laflamme
condition (\ref{EqKnillLaflamme})) says that no information whatsoever
concerning the input state reaches the ancilla.
Mathematically, the Knill-Laflamme condition says that the range of the
conjugate channel lies entirely in the center $\CC\cdot\one_\L$ of $\B(\L)$.
Let us emphasize again that if the space $C$ satisfies the conditions
(\ref{EqKnillLaflamme}) for a map $\Psi$ represented
by a particular set of the Kraus operators $\{A_i\}_{i=1}^m$,
then $C$ also satisfies them for any other set
of Kraus operators $\{B_i\}_{i=1}^{m'}$,
used to represent the same map $\Psi$.

\noindent
Note also that the set of  conditions (\ref{EqKnillLaflamme}),
which express complete darkness,
naturally defines a state $\alpha$,  on the ancilla by a relation
\begin{equation}
\label{EqStateAncilla}
   \Gamma\circ\Psi(\one\otimes Y)=\tr(\a Y)\cdot\one_\L\;.
\end{equation}
satisfied by any $Y$.
This quantum state acting on an auxiliary system is called the
{\it error correction matrix},
since  the density matrix $\alpha_{ij}$ appears in eq. (\ref{EqKnillLaflamme}).
Observe that the density operator $\alpha$ depends only on the map $\Psi$
and not on the concrete form of the Kraus operators $A_i$,
which represent the map and 
determine the matrix representation $\alpha_{ij}$ of $\alpha$.
Relations between matrix elements of the same state
represented in two different basis are governed 
by the Schr{\"o}dinger lemma \cite{BZ06},
also called GHJW lemma \cite{Gi89, HJW93}.

We are now going to prove the equivalence of protection and darkness.
In the case of strong protection and complete darkness this reproduces
and puts into perspective the result of Knill and Laflamme \cite{KL97}
In that case, if the state $\alpha$ is pure, then the decoding operation
$\Delta$ can be realized by a unitary evolution,
Hence the purity constraint for the error correction matrix,
$\alpha=\alpha^2$,
is the correct condition for a decoherence free subspace \cite{LBW99} -- 
see also the proof of Theorem \ref{TheoremProtDark}.
As a quantitative measure, which characterizes to what extent
a given protected space is close to a decoherence free space,
one can use the von Neumann entropy of this state,
$S=-{\rm Tr}\alpha \ln \alpha$. 
This {\em code entropy} \cite{KPZ08} is equal to zero if the 
protected space is decoherence free
or if the information lost can be recovered by a reversible unitary operation. 
Observe that the code entropy $S$ characterizes the map $\Psi$ and 
the code space $C$, but does not depend on the 
particular Kraus form used to represent $\Psi$.

\begin{figure}[htbp]
\centering
\includegraphics[width=0.40\textwidth]{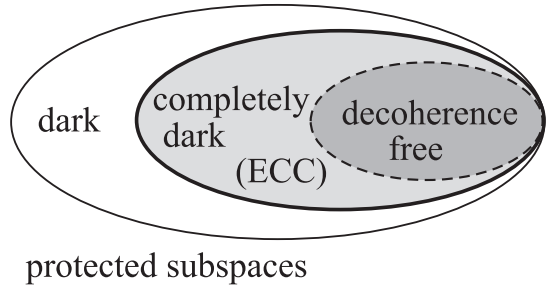}
\caption{Sketch of the hierarchy of protected subspaces.}
\label{fig:fig6k}
\end{figure}

In this way we have determined a hierarchy in the set of protected spaces.
Every decoherence free subspace belongs to the class of 
completely dark subspaces, which correspond to error correction codes.
In turn the completely dark subspaces
form a subset of the set of dark subspaces -- see Fig. \ref{fig:fig6k}.

\medskip

\begin{theorem}{\bf(Equivalence of Protection and Darkness)}
\label{TheoremProtDark}\noindent
Let $\H$, $\M$, and $\L$ be finite dimensional Hilbert spaces.
Let $C:\L\to\H$ and $V:\H\to\H\otimes\M$ be isometries,
and let $\Phi$, $\Psi$ and $\Psi_f$ be as defined 
in (\ref{EqStinespring}), (\ref{EqPsiWide}) and (\ref{EqInstrument}).
Then $C\L$ is weakly protected against the instrument $\Psi_f$ if and only if
$C\L$ is dark for $\Psi_f$.
It is strongly protected against $\Phi$ if and only if it is completely
dark for $\Phi$.
\end{theorem}


\smallskip\noindent{\bf Proof:}


First assume that $C\L$ is strongly protected against $\Phi$,
i.e. (\ref{EqProtected}) holds for some decoding operation $\Delta$.
Let $\Phi(X)=\Psi(X\otimes\one)$ for some compression $\Psi$.
Define
   $$\Theta:\B(\L)\otimes\B(\M)\to\B(\L):
            X\otimes Y\mapsto\Gamma\circ\Psi(\Delta(X)\otimes Y)\;.$$
Then $\Theta(X\otimes\one)=X$ for all $X\in\B(\L)$,
and by Corollary \ref{HeisCor}, since $\Delta(\one)=\one$,
   $$\Gamma\circ\Psi(\one\otimes Y)=\Theta(\one\otimes Y)\in\CC\cdot\one\;.$$
so (\ref{EqComplDark}) holds, and $C\L$ is completely dark for $\Phi$.



\noindent
Conversely, suppose that $C\L$ is completely dark for $\Psi$,
and let $\alpha$ denote the density matrix given by (\ref{EqStateAncilla})
Then we may diagonalize:
   $$\tr(\alpha Y)=\som i a_i\inp{f_i}{Yf_i}$$
for some orthonormal set $(f_i)_{i=1}^{m'}$ (with $m'\le m$)
of $\B(\M)$ and positive numbers $a_1,a_2,\ldots,a_{m'}$ summing up to 1.
Now let
$A_i:=I_{f_i}V$.
Then for all $\psi\in\L$:
\begin{eqnarray*}
   \inp{A_iC\psi}{A_jC\psi}
     &=&\inp{I_{f_i}VC\psi}{I_{f_j}VC\psi}\\
     &=&\inp\psi{C^\dagger V^\dagger(\one\otimes\Ketbra{f_i}{f_j})VC\psi}\\
     &=&\alpha(\Ketbra{f_i}{f_j})\cdot\|\psi\|^2\\
     &=&a_i\delta_{ij}\cdot\|\psi\|^2\;.
\end{eqnarray*}
So the ranges of $A_iC$ and $A_jC$ are orthogonal for $i\ne j$ and 
$A_i$ is homometric on $C\L$.
Now define $D_i$ for $i=1,2,\ldots,m'$
on these orthogonal ranges by
   $$D_i\ph=0\quad\hbox{if}\quad\ph\perp\hbox{Range}\,(A_iC),\qquad
     D_i A_i C\psi=\sqrt{a_i}\psi\;.$$
($D_i$ ``rotates back'' the action of $A_iC$.)
Let $\Delta$ denote the operation
   $$\Delta(Z):=\somp i D_i^\dagger Z D_i
                +\rho(Z)\left(\one_\H-\somp j D^\dagger D_j\right)\;.$$
for some arbitrary state $\rho$ on $\B(\L)$.
(The term with $\rho$ is intended to ensure that
$\Delta(\one_\L)=\one_\H$.)
Then we have for all $Z\in\B(\L)$:
\begin{eqnarray*}
\Gamma\circ\Phi\circ\Delta(Z)
   &=&\somp j \somp i C^\dagger A_j^\dagger D_i^\dagger Z D_i A_j C\\
   &=&\somp j \somp i  \frac{1}{a_i}C^\dagger A_j^\dagger A_i CZ
                                 C^\dagger A_i^\dagger A_j C
    = \somp {ij} \delta_{ij}a_i Z
    = Z\;.
\end{eqnarray*}
So $C\L$ is strongly protected against $\Phi$ by (\ref{EqProtected}).


Now let us prove the equivalence between weak protection and darkness.
Assume that $C\L$ is weakly protected against $\Psi_f$,
i.e. (\ref{EqWeakProt}) holds for some $\Delta_f:\B(\L)\to\B(\H)\otimes\CC^m$,
say $\Delta_f(X)=\sum_{j=1}^m \Delta^j_f(X)\otimes e_j$.
Define $\Theta:\B(\H)\otimes\CC^m\to\B(\H)$ by
   $$\Theta(X\otimes g)
:=\sum_{j=1}^m g(j)\Gamma\circ\Psi_f(\Delta^j(X)\otimes e_j)\;.$$
Then by (\ref{EqWeakProt}), $\Theta(X\otimes\one)=X$ for all $X\in\B(\L)$.
Hence by Corollary \ref{HeisCor},
   $$\Gamma\circ\Psi_f(\one\otimes e_i)=\Theta(\one\otimes e_i)
      \in\B(\H)'=\CC\cdot\one_\L\;.$$
So (\ref{EqDark}) holds, and $C\L$ is dark for $\Psi_f$.


Conversely, assuming that $C\L$ is dark for $\Psi_f$,
then $A_lC$ is homometric on $\L$ by (\ref{EqDarkKraus}),
and we may define $D_l:\H\to\L$
by
   $$D_l A_l C\psi:=\sqrt{\lambda_l}\psi\quad\hbox{if } \psi\in\L,
     \qquad D_l\ph=0 \quad\hbox{ if }\ph\perp\hbox{Range}\,(A_lC)\;.$$
(Briefly: $D_l=C^\dagger A_l^\dagger/\sqrt{\lambda_l}$ if $\lambda_l\ne0$,
zero otherwise.)
Define the decoding operation $\Delta_f:\B(\L)\to\B(\H)\otimes\CC^m$ by
$$\Delta_f(Z):=\bigoplus_{l=1}^m
            \left(D_l^\dagger Z D_l+(\one_\H-D_l^\dagger D_l)\rho(Z)\right)$$
for some arbitrary state $\rho$ on $\B(\L)$.
Then, for $Z\in\B(\L)$:
\begin{eqnarray*}
\Gamma\circ\Psi_f\circ\Delta_f(Z)
    &=&\Gamma\circ\Psi_f
       \left(\som l \bigl(D_l^\dagger Z D_l+(\one-D_l^\dagger D_l)\rho(Z)\bigr)
               \otimes e_l\right)\\
    &=&C^\dagger V^\dagger
       \left(\som l \bigl(D_l^\dagger Z D_l+(\one-D_l^\dagger D_l)\rho(Z)\bigr)
               \otimes\ketbra{f_l}\right)VC\\
    &=&\som l C^\dagger A_l^\dagger D_l^\dagger Z D_l A_l C
     =\som l \frac{1}{\lambda_l}(C^\dagger A_l^\dagger A_l C)
                                    Z(C^\dagger A_l^\dagger A_l C)
     =\som l \lambda_l Z = Z\;.
\end{eqnarray*}
{\hfill $\square$}

\section{Compression problems and generalized numerical range}
\label{sec:comp}
For a given channel $\Phi:\B(\H)\to\B(\H)$ we are interested in the 
protected subspaces of $\H$.
These are the subspaces on which the compressions of $A_i^\dagger A_j$
act as scalars.
In this section we review this compression problem.

\noindent
Let $T$ be an operator acting on a Hilbert space $\H$ of dimension $n$, say.
For any $k\geq 1$, define the
{\it rank-$k$ numerical range} of $T$ to be the subset of the
complex plane given by
\begin{eqnarray}
\label{nrdefn}
\Lambda_k(T) = \big\{ \lambda\in {\CC} : C^\dagger T C = \lambda\one
\,\, {\rm for \,\, some \,\,} C:\CC^k\to\H\big\},
\end{eqnarray}
The  elements of $\Lambda_k(T)$ 
can be called ``compression-values''
for $T$, as they are obtained through compressions of
$T$ to a $k$-dimensional {\sl compression subspace}. 
The case $k=1$ yields the standard 
numerical range for operators \cite{Bh97} 
\begin{eqnarray}
\Lambda_1(T) = \{ \langle \psi| {T\psi}\rangle \,
: \,  | \psi\rangle \in {\cal H} \,  ,  \langle \psi | \psi \rangle  = 1 \}.
\end{eqnarray}

\noindent
It is clear that
\begin{eqnarray}\label{inclusions}
\Lambda_1(T) \supseteq \Lambda_2(T) \supseteq \ldots
\supseteq \Lambda_n(T).
\end{eqnarray}

The sets $\Lambda_k(T)$, $k>1$, are called 
{\it higher-rank numerical ranges} \cite{CKZ06b,CHKZ07}.
For any normal operator acting on ${\cal H}_n$ this is a compact subset
of the complex plane.
For unitary operators this set is
included inside every convex hull $({\rm co}\, \Gamma)$, where $\Gamma$ is
an arbitrary $(n+1-k)$-point subset (counting multiplicities) of the
spectrum of $T$ \cite{CKZ06b}.
It was recently shown that for any normal operator the sets  $\Lambda_k(T)$ 
are convex \cite{LS08,Wo08}
while further properties of higher rank numerical range
were investigated in \cite{CGHK08,LPS07,LPS07b}.

\noindent
The higher rank numerical range is easy to find
for any  Hermitian operator,  $T=T^{\dagger}$
acting on an $n$-dimensional Hilbert space $\H$.
Let us quote here a useful result proved in  \cite{CKZ06b}.

\medskip

\begin{lemma}
\label{LemNumRange}
Let $x_1 \le x_2 \le \dots \le x_n$ denote the ordered
spectrum (counting multiplicities) of  a hermitian operator  $T$.
The rank-$k$ numerical range of $T$ is given by the interval
\begin{equation}
\Lambda_k (T) =[x_k, x_{n+1-k} ] \ ,
\label{herm1}
\end{equation}
\end{lemma}

\begin{figure}[htbp]
\centering
\includegraphics[width=0.75\textwidth]{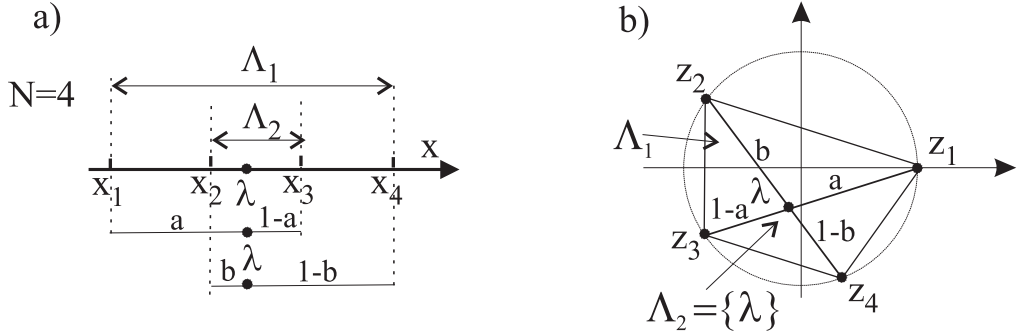}
\caption{Standard numerical range $\Lambda_1$ 
and higher rank numerical range $\Lambda_{2}$
for a)  Hermitian operator $T$ of size $4$ and
b)  non--degenerate unitary $U\in U(4)$.
Observe similarity in finding the weights $a$ and $b$
used to construct superposition of states forming the subspace $C\L$
in both problems.}
\label{fig:fig1}
\end{figure}

\medskip\noindent
Note that the higher rank numerical range of a hermitian $T$ is nonempty for 
any $k\le {\rm int}[(n+1)/2]$.
Let us demonstrate an explicit construction
of a compression to $\CC^2$ which solves 
equation (\ref{nrdefn}) for a Hermitian matrix $T$
of size $n=4$. The latter's eigenvalue equation reads
$T |\phi_i\rangle = x_i |\phi_i\rangle$.
Choose any real $\lambda \in \Lambda_2(T)=[x_2,x_3]$.
It may be represented as a convex combination of
two pairs of eigenvalues $\{x_1, x_3\}$ and $\{x_2,x_4\}$
-- see Fig. \ref{fig:fig1}a.
Writing 
\begin{equation}
\lambda \ = \ (1-a) x_1+ a x_3 \ = \ (1-b) x_2+ b x_4
\label{lambd1}
\end{equation}
one obtains the weights
\begin{equation}
\label{cosine}
a  = \frac{\lambda - x_1}{x_3-x_1} =: \sin^2\theta_1 
{\rm \quad and \quad}
b  = \frac{\lambda - x_2}{x_4-x_2} =: \sin^2\theta_2 
\end{equation}
which determine real phases  $\theta_1$ and $\theta_2$. 
These phases  allow us to define an isometry $C:\CC^2\to\H$ by 
\begin{eqnarray}
\label{rank2}
C:\left\{ \begin{array}{ccl}
        e_1&\mapsto& \cos\theta_1|\phi_1\rangle+\sin\theta_1|\phi_3\rangle  \\
        e_2&\mapsto& \cos\theta_2|\phi_2\rangle+\sin\theta_2 |\phi_4\rangle 
\end{array}\right. , 
\end{eqnarray}
Observe that
\begin{equation}
\inp{e_1}{C^\dagger T Ce_1}=\cos\theta_1 x_1
\langle \phi_1| \psi_1\rangle  + \sin\theta_1 x_3
\langle \phi_3 | \psi_1\rangle  = (1-a) x_1 +  a x_3 = \lambda.
\end{equation}
Similarly, we have $\inp{e_2}{C^\dagger T C e_2}= \lambda$.
Further, we also have
$\inp{e_1}{C^\dagger T C e_2}= 0 = \inp{e_2}{C^\dagger T C e_1}$.
It follows that $C^\dagger T C=\one$,
and the isometry (\ref{rank2}) provides a solution of 
the compression problem  (\ref{nrdefn}) as claimed.
Note that one can select  another pairing of eigenvalues,
and the choice  $\{x_1, x_4\}$ and $\{x_2,x_3\}$
allows us to get in this way 
another subspace $C'\L$ spanned by vectors
obtained by a superposition of states 
$|\phi_1\rangle $ with $|\phi_4\rangle$ and
$|\phi_2\rangle $ with $|\phi_3\rangle$
respectively.

For a given operator $T$ one may try to solve its compression
equation (\ref{nrdefn}) and  look for its numerical range 
$\Lambda_k(T)$. Alternatively, one may be interested 
in the following simple {\sl compression problem}:
For a given operator $T$ 
find all possible subspaces $C\L$ 
of a fixed size $k$ which satisfy (\ref{nrdefn}). 

Furthermore, it is natural to raise a more general,
{\sl joint compression problem} of order $M$.
For a given set of $M$ operators $\{T_1, \dots T_M\}$
acting on $\H_n$ find a subspace $C\L$ of dimensionality $k$
which solves simultaneously $M$ compression problems:
\begin{equation}
C^\dagger T_m C=\lambda_m\one
{\rm \quad for \quad}
m=1,\dots,M \ .
\label{jointM}
\end{equation}
Note that all compression constants, $\lambda_m \in \Lambda_k(T_m)$,
can be different, but the isometry $C$ needs to be the same.

\section{Dark subspaces}
In this section we provide several results
concerning existence of darks spaces for 
several classes of quantum maps.

\subsection{Random external fields}
Consider a noisy channel $\Phi$ given by
\begin{equation}
\Phi_U(X)= \sum_{i=1}^r q_i \, U_i^\dagger X U_i , 
\label{REF}
\end{equation}
where all operators $U_i$ are unitary while positive weights $q_i$ 
sum up to unity. 
Such maps are called {\sl random external fields} \cite{AL87}
or random unitary channels. The standard Kraus form  (\ref{EqDecomp})
is obtained by setting $A_i=\sqrt{q_i} U_i$.

In this Kraus decomposition the whole space, and hence every subspace, is dark.
This corresponds to the fact that the choice between the unitaries,
which is made with the probability distribution $(q_1,\ldots,q_r)$,
gives no information on the quantum state.
And indeed, knowledge of the ``external field'',
i.e. of the outcome $i$, permits us to undo,
by the inverse of $U_i$, the action of the channel.

\subsection{Rank two quantum channels}

Let us now analyze a rank two channel,  
\begin{equation}
\rho'=\Phi_2(\rho)=  A_1 \rho A_1^{\dagger} +  A_2 \rho A_2^{\dagger} \ ,
\label{rank2b}
\end{equation}

\begin{lemma}
For any Kraus representation of
any rank-two channel acting on a system of size $N$ 
there exist a dark subspace of dimension $k={\rm int}[(N+1)/2].$
\end{lemma}

{\bf Proof.}
We need to solve a joint compression problem (\ref{jointM})
of order two, for two Hermitian operators 
$T_1=A_1^{\dagger} A_1$ and $T_2=A_2^{\dagger} A_2$.
Due to Lemma \ref{LemNumRange} there exists a subspace $P_k$ 
of dimension $k={\rm int}[(N+1)/2]$ which solves the compression problem
for the Hermitian operator $T_1$ of size $N$.
It is also a solution of the compression problem for the
other operator, since the trace preserving condition implies
$T_2={\mathbbm 1}-T_1$. $\square$ 

\subsection{Biased permutation channel}

Consider a quantum map acting on a system of arbitrary size $n$
described by the Kraus form (\ref{EqDecomp}). Let us assume that
all Kraus operators are given by 'biased permutations'
\begin{equation}
A_i = P_i \sqrt{D_i} \ ,  \quad i=1,\dots, r \ .
\label{biased}
\end{equation}
where $D_i$ is a diagonal matrix containing non-negative entries,
and $P_i$ denotes an arbitrary permutation of the $N$-element set.
Hence all elements of the POVM form diagonal matrices, 
\begin{equation}
T_i=A_i^{\dagger} A_i  =  \sqrt{D_i} P_i^{\dagger} P_i \sqrt{D_i} =  D_i \ ,
\label{biased1}
\end{equation}
in general not proportional to identity.
Note that the Kraus operators defined in this way need not to be Hermitian.
To satisfy the trace preserving condition (\ref{EqIsometry})
we need to assume that $\sum_{i=1}^r D_i={\mathbbm 1}$.
Let us define an auxiliary rectangular matrix of size $r \times N$,
namely $S_{im}:=(D_i)_{mm}\ge 0$.
Then  the above constraints for the matrices $D_i$ is equivalent to the
 statement that $S$ is {\sl stochastic}, since the sum of all elements in each
column is equal to 1,
\begin{equation}
\sum_{i=1}^r S_{im}= 1
{\rm \quad for \quad}
m=1,\dots, N \ .
\label{biased2}
\end{equation}
A map described by Kraus operators fulfilling
relations (\ref{biased}) and (\ref{biased2})
will be called a {\sl biased permutation channel}.

\medskip

We are going to construct a dark space for a wide class
of such channels. For simplicity assume that
the size of the system is even, $N=2k$.
Let us additionally assume that 
all elements in each row of $B$ are ordered
(increasingly or decreasingly) 
and that the matrix $S$
enjoys a symmetry relation,
\begin{equation}
 S_{i,m}+S_{i,n-m+1}={\rm const} =: \lambda_i
{\rm \quad for \quad}
i=1,\dots,r; \quad m=1,\dots, k=n/2 \ .
\label{biased3}
\end{equation}
Then the numbers $\lambda_i$ can be defined by a sum
of the entries in each row, $\lambda_i=\frac{2}{N}  \sum_{m=1}^N S_{im}$.

\medskip

\begin{lemma}
Assume that a biased permutation channel acting
 on a system of size $N=2k$  possesses the symmetry relation (\ref{biased3}).
 Then it has a dark space of dimension $k=n/2$.  
\end{lemma}

{\bf Proof.}
We need to find a joint compression subspace for the set of $r$
elements of POVM given by diagonal matrices $D_i$, with $i=1,\dots, r$.
Since these matrices commute, they have the same set of eigenvectors, 
denoted by $|v_m\rangle$, $m=1, \dots, N$. 
Due to symmetry relation (\ref{biased3}) we know that
the barycenter of each spectrum,   $\lambda_i $
belongs to the higher rank numerical range, $\Lambda_k(D_i)$.
Furthermore, this relation shows that (for any $i$)
the number $\lambda_i$ can be represented as a sum of two eigenvalues of $D_i$
with the same weights, $\lambda_i=\frac{1}{2} (D_i)_{mm}
 + \frac{1}{2} (D_i)_{m'm'}$ with $m'=n+1-m$. By construction
this property holds for all operators $D_i$, $i=1,\dots r$.
Hence  the general construction of the higher order numerical
range for Hermitian operators \cite{CKZ06} implies
that the subspace 
\begin{equation}
C_k:= \sum_{i=1}^k |\psi_i\rangle \langle \psi_i|
{\rm \quad where \quad}
|\psi_i\rangle := \frac{1}{\sqrt{2}} (|v_i\rangle +|v_{1-i+N} \rangle ) 
\label{biased4}
\end{equation}
fulfills the joint compression problem for all operators
$T_i=D_i$, $i=1,\dots r$. Hence this subspace is dark as advertised.  $\square$ 

\medskip

To watch the above construction in action consider
a three biased permutation channel acting on a two qubit system.
Hence we set $r=3$ and $N=4$, and 
assume that five real  weights satisfy
$0<a<b<x/2<1/2$ and $0<c<d<x/2$.
They can be used to define the channel
by a stochastic matrix $S$
\begin{equation}
 S=\begin{pmatrix}
          a  &  b & x-b & x-a \\
          c  &  d& x-d  & x-c \\
         a' & b' & b'' & a'' \\
 \end{pmatrix} \ ,
\label{matB}
\end{equation}
where $a'=1-a-c$, $b'=1-b-d$, $a''=1-2x+a+c$ and $b''=1-2x+b+d$.
Note that this matrix satisfies the symmetry condition (\ref{biased3}),
the elements in each row are ordered,
 while mean weights in each row read $\lambda_1=\lambda_2=x/2$ and
$\lambda_3=2(1-x)$.

To complete the definition of the channel we need to specify three
permutation matrice of size four.
For instance let us choose $P_1=P_{(1,2,3,4)}$, $P_2=P_{(1,2),(3,4)}$
and $P_3=P_{(1,4,3,2)}$, where according to the standard notion,
the subscripts contain the permutation cycles.
Then the biased permutation channel is defined by the
three Kraus operators 
\begin{equation}
 A_1=\begin{pmatrix}
          0  &    \sqrt{b} & 0 & 0 \\
          0  &           0 & \sqrt{x-b} & 0 \\
         0    &   0     &  0 &    \sqrt{x-a} \\
        \sqrt{a}   &   0     &   0   &  0 \\
 \end{pmatrix} \ ,  \ 
 A_2=\begin{pmatrix}
          0  &    \sqrt{d} & 0 & 0 \\
      \sqrt{c}&           0 & 0 & 0 \\
         0    &   0     &  0 &    \sqrt{x-c} \\
        0    &   0     &   \sqrt{x-d} &  0 \\
 \end{pmatrix} \ , \ 
A_3=\begin{pmatrix}
          0  &   0  & 0 & \sqrt{a''}  \\
           \sqrt{a'}  &           0 & 0 & 0 \\
         0    &    \sqrt{b'}    &  0 &    0 \\
         0  &   0     &   \sqrt{b''}  &  0 \\
 \end{pmatrix} \ ,
\end{equation}
which satisfy the trace preserving condition (\ref{EqIsometry}).

Since the barycenter $\lambda_i$ of the spectrum
of the POVM element $T_i=D_i$ (given by a row of matrix (\ref{matB})),
is placed symmetrically, in all three cases
it can be represented by a convex combination of pairs of eigenvalues
with weights equal to $1/2$. Thus we define  two pure states 
\begin{equation}
|\psi_1\rangle := \frac{1}{\sqrt{2}}(|v_1\rangle +|v_4 \rangle)\ , \quad
|\psi_2\rangle := \frac{1}{\sqrt{2}}(|v_2\rangle +|v_3 \rangle)\  ,
\label{alpha}
\end{equation}
and the two dimensional subspace spanned by them,
$C= |\psi_1\rangle \langle \psi_1| + |\psi_2\rangle \langle \psi_2|$.
It is easy to verify that the subspace $C$
satisfies  $C^\dagger T_1 C=\lambda_1\one=C^\dagger T_2 C$ while 
$C^\dagger T_3 C=\lambda_3\one$ so this space is dark.
Note that the subspace $C\L$ cannot be used to design an 
error correcting code since
$C^\dagger A_1^{\dagger}A_2 C \notin\CC\cdot\one$.

\subsection{Composed systems and separable channels}

Consider a bipartite system of size $n=n_A \times n_B$.
A quantum operation $\Phi$ acting on this bipartite system
is called {\sl local}, if it has a tensor product structure,
$\Phi=\Phi_A \otimes \Phi_B$, where both  
maps $\Phi_A$ and $\Phi_B$ are completely positive and
preserve the identity.
If for both individual operations, $\Phi_A$ and $\Phi_B$, 
there exist protected subspaces $C_k$ and $Q_l$ respectively, 
then the product subspace $C_k\otimes Q_l$ of size $kl$
 is also a protected subspace for the composite map $\Phi_A \otimes \Phi_B$.

Similar protected subspaces of the product form can be constructed 
 for a wider class  of {\sl separable maps} (see e.g. \cite{BZ06}),
\begin{equation}
\rho'=\Phi^*(\rho) =\sum_{i=1}^{r} (A_i \otimes B_i)  \rho (A_i\otimes
B_i)^{\dagger} .
 \label{separmap}
\end{equation}

Assume that a subspace $C_k\in {\cal H}_{N_A}$ is a solution of the joint
compression problem  for the set of $r$ operators $A_i^{\dagger} A_i$, 
while a subspace $Q_l \in {\cal H}_{N_B}$ does the job
for the set of $r$ operators $B_i^{\dagger} B_i$.
It is then easy to see that  the product subspace
$C_k\otimes Q_l$ of dimension $kl$ is a dark subspace for the 
separable map  (\ref{separmap}).

It is straightforward to extend lemmas 3 and 4 
for separable maps acting on composite systems
and apply them to construct  protected subspaces 
with a product structure. On the other hand, 
if for certain problems such product code subspace do not exist, 
one may still find a code subspace spanned by entangled states.
Such a problem for the tri--unitary model 
is solved in following section.

\section{Unitary noise and error correction codes}

In this section we are going to study 
multiunitary noise  (\ref{REF}),
also called  random external fields,
and look for existence of error correction codes,
i.e. completely protected subspaces.
In general the number $r$ of unitary operators
defining the channel can be arbitrary
but we will restrict our attention to the cases in which this number is small.

\subsection{Bi--unitary noise model}

The case in which $r=2$, referred to as
{\sl bi-unitary noise} was
recently analyzed in \cite{CKZ06,CHKZ07}.
Let us rewrite the dynamics in the form
\begin{equation}
\rho'=\Phi^*(\rho) =  q V_1^\rho V_1^{\dagger}+ (1-q) V_2 \rho V_2^{\dagger} \ . 
\label{biunitary}
\end{equation}
and assume that we deal with the system of two qubits.
Then both unitary matrices  $V_1$ and $V_2$ belong to $U(4)$
while probability  $p$  belongs to $[0,1]$.
The problem of finding the compression $C$
for the above map is shown to be equivalent to 
the case 
\begin{equation}
\rho'' = \Phi^*(\rho) = q \rho+
  (1- q) U \rho U^{\dagger}
 \label{biunitary2}
\end{equation}
where $U=V_1^{\dagger}V_2$. 

\noindent
Thus the error correction  matrix $\alpha$ of size two 
defined by eq. (\ref{EqStateAncilla}) reads
\begin{equation}
\alpha \ = \ 
\left(\begin{matrix} 
  q                                      &   \sqrt{q(1-q)} \lambda \\ 
 \sqrt{q(1-q)} \lambda^* &   1-q 
\end{matrix}\right)
\label{lambda1}
\end{equation}
where $\lambda$ is solution of the compression problem for $U$
\begin{equation}
C^\dagger U C= \lambda\cdot\one  \ .
\label{comp1}
\end{equation}
Thus to find the error correction space
for the bi--unitary model it is sufficient to solve the 
compression equation for a single operator $U$.
A solution exists for any unitary $U$  \cite{CKZ06}, 
but for simplicity we will consider here the generic case
if  the spectrum of $U$ is not degenerated.
Assume that the phases these unimodular numbers
$z_1,...,z_4$  are ordered and that 
$|\psi_i\rangle$ denote the corresponding eigenvectors.

\noindent
Let $\lambda$ denote the  intersection 
point between two  chords of the unit 
circle,  $z_1z_3$ and $z_2z_4$; compare Fig. \ref{fig:fig1}b.
This point can be represented as a convex combination of each 
pair of complex eigenvalues,
\begin{equation}
\label{decomp2}
\lambda  \ = \  (1-a) z_1 + a z_3  \ = \
                        (1-b) z_2 + b z_4  \  ,
\end{equation}
where the non--negative weights read 
\begin{equation}
\label{cosine2}
a  = \frac{\lambda - z_1}{z_3-z_1} =: \sin^2\theta_1 
{\rm \quad and \quad}
b  = \frac{\lambda - z_2}{z_4-z_2} =: \sin^2\theta_2 
\end{equation}
and determine real phases  $\theta_1$ and $\theta_2$. 
Note similarity with respect to the construction 
used in the Hermitian case, 
in which  (\ref{lambd1}) represents a 
convex combination of points on the real axis.
In an analogy with the reasoning performed for a hermitian $T$
we define according to (\ref{rank2}) an orthonormal pair of vectors 
$| \psi_1\rangle$ and  $| \psi_2\rangle$
and define the associated isometry $C:e_j\mapsto\psi_j$.
Since 
$ \langle U\psi_1 |\psi_1\rangle =  \lambda =
\langle U\psi_2 | \psi_2\rangle $
and  
$\langle U\psi_1 |\psi_2\rangle = 0 = \langle U \psi_2| \psi_1\rangle $
then  $C U C = \lambda {\mathbbm 1}$. 
Therefore  $\lambda$ belongs to
$\Lambda_2(U)$ as claimed and the range of $C$
provides the error correction code for 
the bi-unitray noise (\ref{biunitary2}) acting on a 
two-qubit system.

In the case of doubly degenerated spectrum of $U$
the complex number $\lambda$ is equal to the degenerated eigenvalue,
so its radius, $|\lambda|$, is equal to unity.
In this case the matrix $\alpha$ given in (\ref{EqKnillLaflamme}) 
represents a pure state, $\alpha=\alpha^2$, so the two--dimensional
subspace spanned by both eigenvectors corresponding
to the degenerated eigenvalues is {\sl decoherence free}.
 
Bi--unitary noise model for higher dimensional  systems
was analyzed in \cite{CHKZ07}. It was shown in this work 
that for a generic $U$ of size $N$
there exists a code subspace of dimensionality 
$k={\rm int}[(N+2)/3]$.
This result implies that for a system of $m$ qubits
and a generic $U$
of size $N=2^m$ there exists an error correction code
supported on $m-2$ qubits.
Furthermore, if $N=d^m$ and $d\ge 3$,
there exists a code supported on $m-1$
quantum  systems of size $d$.

\subsection{Tri--unitary noise model}
\label{sec:tri} 

Consider now a model of noise described by three unitary operations
acting on a bipartite, $N=2 \times N_B$ system,
\begin{equation}
\rho'=\Phi^*(\rho) =  q_1 V_1 \rho V_1^{\dagger}+ q_2 V_2 \rho W_2^{\dagger} +
(1-q_1-q_2)   V_3 \rho V_3^{\dagger} \ .
 \label{triunitary}
\end{equation}
Performing a unitary rotation  in analogy to (\ref{biunitary2})
we obtain an  equivalent form
\begin{equation}
\rho''  =\Phi^*(\rho) =  q_1 \rho  + q_2 U_1\rho U_1^{\dagger}+ (1-q_1-q_2) U_2
\rho
U_2^{\dagger} \ . 
 \label{triunitary2}
\end{equation}
The model is thus characterized by 
 two unitary matrices of size $N$, namely 
 $U_1=V_1^{\dagger} V_2$ and $U_2=V_1^{\dagger}V_3$. 
and two weights $q_1$ and $q_2$,
which we assume to be positive
with their sum smaller than unity.

To find a simplest error correction code
for this model one needs to find a two-dimensional subspace,  
which forms a joint solution of three compression problems
\begin{equation}
\left\{
\begin{array}{ccc}
C^\dagger U_{1}C &=& \lambda_{U_{1}}\one\\
C^\dagger U_{2}C &=& \lambda_{U_{2}}\one\\
C^\dagger W C &=& \lambda_{W}\one\\
\end{array}\right.,
\label{eq:rkompresji}
\end{equation}
where $W=U_{1}^{\dagger}U_{2}$.
Each of the above three problems may be solved
using the notion of the higher rank numerical range of a unitary matrix.
However, for generic unitary matrices 
$U_1$ and $U_2$ of size $4$ the
corresponding  compression subspaces do differ.
Thus for a typical choice of the unitary matrices
the tri--unitary noise model will not have an error correction code,
for which it is required that the subspace $C$ solves all three 
problems simultaneously. 

\begin{figure}[htbp]
\centering
\includegraphics[width=0.35\textwidth]{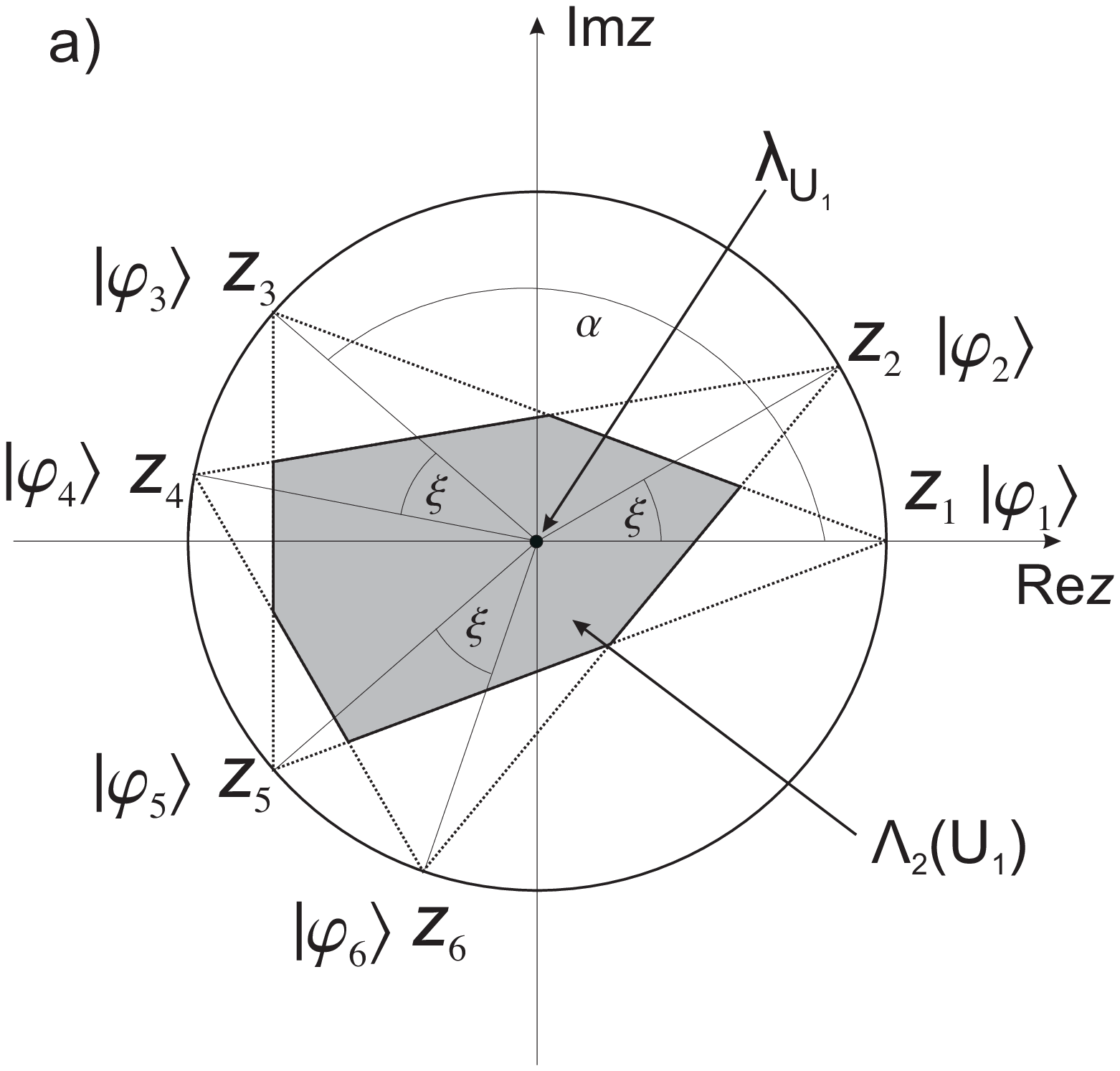}
\includegraphics[width=0.35\textwidth]{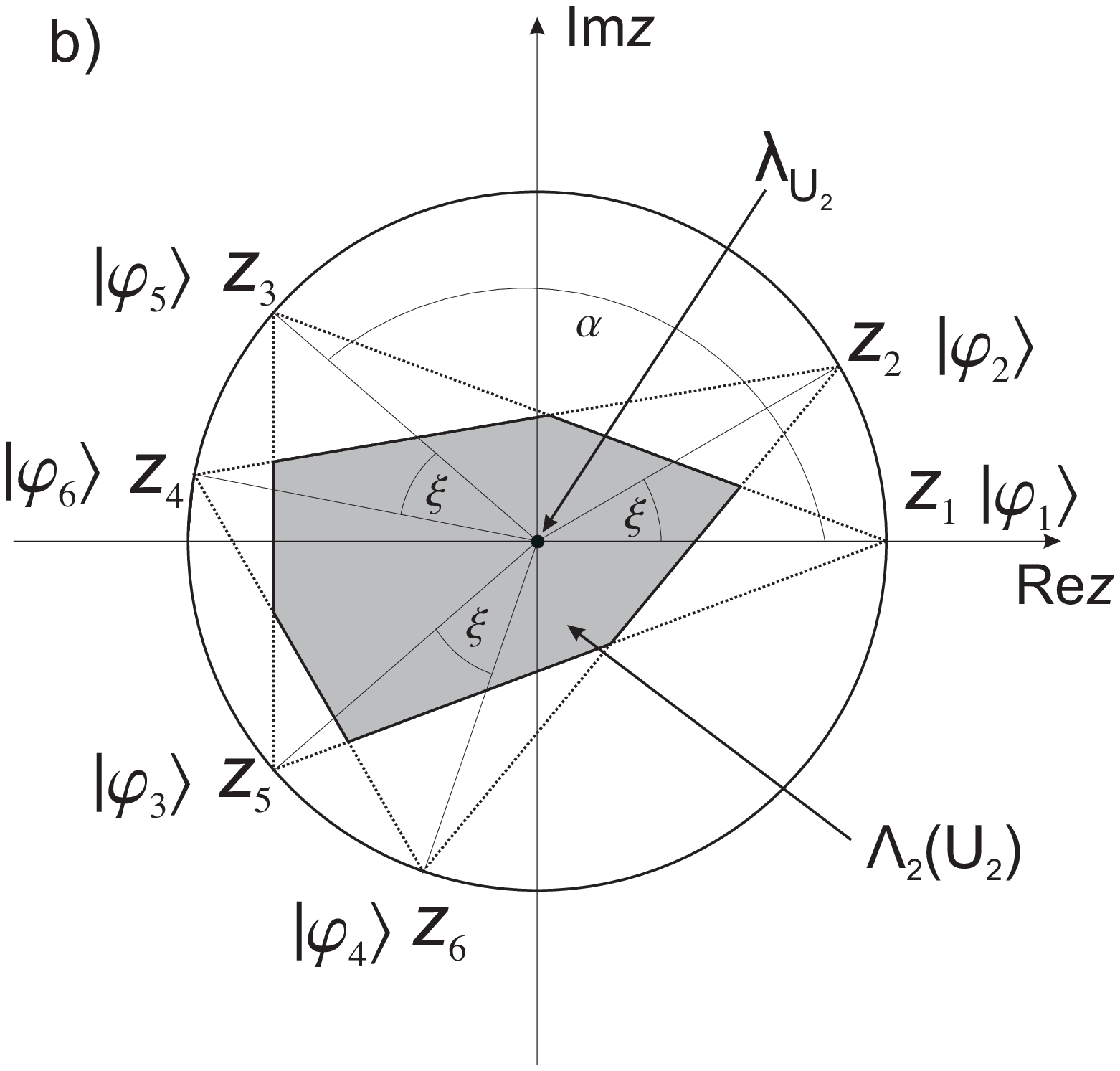}
\caption{Numerical range (gray space): a) $\Lambda_{2}(U_{1})$; b)
$\Lambda_{2}(U_{2})$}
\label{fig:TUCUV}
\end{figure}

There exist several examples of two commuting
matrices $U_1$ and $U_2$ of size $N=4$, 
such that they possess the same solution
$C$ of the compression problem.
However, to assure that it coincides with the
solution of the same problem for  $W=U_{1}^{\dagger}U_{2}$, 
we will analyze an exemplary system of size  $n=2 \times 3$.
Consider two unitary matrices of a tensor product form,
\begin{equation}
\left\{
\begin{array}{ccc}
U_{1}&=&U_{A}^{\dagger}\otimes U_{B}\\
U_{2}&=&U_{A}\otimes U_{B}
\end{array}\right.
\label{eq:rtens}
\end{equation}
where 
\begin{equation}
U_{A}=\left(
\begin{array}{ccc}
1 & 0 & 0\\
0 & e^{-i\alpha} & 0\\
0 & 0 & e^{i\alpha}
\end{array}\right)
{\quad \rm and \quad}
U_{B}=\left(
\begin{array}{cc}
1 & 0\\
0 & e^{i\xi}
\end{array}\right) \ .
\end{equation}
Observe that $U_1$ and $U_2$ do commute,
so they share the same set of eigenvectors.
Assume that the phases satisfy 
$\alpha \in (\pi/2,\pi)$
and $\xi \in \bigl( 0, {\rm min} \{\alpha,2(\pi - \alpha) \} \bigr)$.
Then the ordered spectra of both matrices read
\begin{equation}
U_{1}={\rm diag} \Big\{ 1, e^{i \xi} , e^{i\alpha},  e^{i(\alpha+\xi)}, 
e^{-i\alpha}, e^{i(\xi-\alpha)} \Big\} ,
{\quad  \quad}
U_{2}={\rm diag} \Big\{ 1, e^{i \xi} , e^{-i\alpha},  e^{i(\xi-\alpha)}, 
e^{i\alpha}, e^{i(\alpha+\xi)} \Big\} ,
\label{daiagU}
\end{equation}
and differ only by the order of the eigenvalues. 
Both unitary matrices are represented in Fig.  \ref{fig:TUCUV}
in which $z_i$, $i=1,\dots,6$ denote the ordered eigenvalues
of $U_1$ while  $|\varphi_{i}\rangle, i=1,\ldots,6$ 
are eigenvectors of this matrix.
The same states form also the set of eigenvectors 
of $U_2$, but they correspond to other eigenvalues.
Let $z_i'$ denote the ordered eigenvalues of $U_2$.
Then  $|\varphi_3\rangle$ corresponds to $z_3'=z_5$ 
while $|\varphi_5\rangle$ corresponds to $z_5'=z_3$.

\begin{figure}[htbp]
\centering
\includegraphics[width=0.35\textwidth]{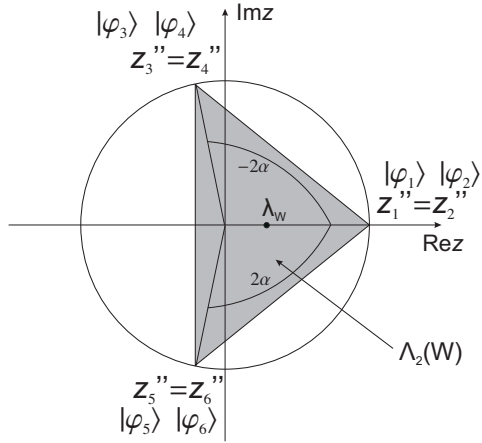}
\caption{Numerical range $\Lambda_{2}(W)$ is represented by 
a dark triangle}
\label{fig:TUCW}
\end{figure}

The third of the unitaries also has also a tensor product form, 
\begin{equation}
\label{eq:iltens1}
W=U_{1}^{\dagger}U_{2}=(U_{A}^{\dagger}\otimes U_{B})^{\dagger}(U_{A}\otimes
U_{B}) =U_{A}^{2}\otimes \mathbbm{1}_{2}.
\end{equation}
Hence the spectrum of $W$, denoted by $z_i''$,
consists of three pairs of doubly degenerated eigenvalues, 
$ W={\rm diag} 
\Big\{1, 1, e^{-2i\alpha}, e^{-2i\alpha}, e^{2i\alpha}, e^{2i\alpha} \Big\}$, 
see Fig.  \ref{fig:TUCW}.

Numerical range of rank two for matrices $U_1$, $U_2$ and $W$
is shown in the pictured as a gray region.
Each point $\lambda\in \Lambda_2(U_1)$
offers a subspace $C_2$ which forms a solution
of the first of three equations (\ref{eq:rkompresji}).
However, the other two equations restrict 
 further constraints for $\lambda$.

To construct an error correction code for the
tri-unitary noise model we are going to follow the strategy
used above for solving the compression problem:
we split the Hilbert space into a direct
sum of two subspaces of size three, 
and try to construct a single state in each subspace. 
More formally we define the subspace 
\begin{equation}
C_{2}=\sum_{i=1}^{2}|\psi_{i}\rangle\langle\psi_{i}|
\label{eq:p2}
\end{equation}
where each state is obtained by a coherent superposition of three eigenstates
of $U_1$,
\begin{equation}
\left\{
\begin{array}{ccc}
|\psi_{1}\rangle&=&\sqrt{a_{1}}|\varphi_{1}\rangle+\sqrt{a_{3}}|\varphi_{3}\rangle+\sqrt{a_{5}}|\varphi_{5}\rangle\\
|\psi_{2}\rangle&=&\sqrt{a_{2}}|\varphi_{2}\rangle+\sqrt{a_{4}}|\varphi_{4}\rangle+\sqrt{a_{6}}|\varphi_{6}\rangle
\end{array}\right..
\label{eq:wektfalowe}
\end{equation}
Since the unitary operators $U_i$ can be expressed as 
 tensor product of diagonal matrices (e.g. $U_{2}=U_{A} \otimes U_{B}$),
 their joint set of eigenvectors consits of product pure states only.
On the other hand, the states $|\psi_1\rangle$ and $|\psi_2\rangle$
are by construction entangled. 

The weights $a_1$ are defined as a weights obtained by representing 
point $\lambda$ by a convex combination of the triples of eigenvalues.
Since we wish to get a space $C$ being a joint solution of all three equations
(\ref{eq:rkompresji}), we are going to require that
the same weights $a_i$ can be used to form the compression value $\lambda$
as a combination of both triples of eigenvalues for each spectrum, 
\begin{equation}
\left\{
\begin{array}{ccccc}
\lambda_{U_{1}}&=&a_{1}z_{1}+a_{3}z_{3}+a_{5}z_{5}&=&a_{2}z_{2}+a_{4}z_{4}+a_{6}z_{6}\\
\lambda_{U_{2}}&=&a_{1}z'_{1}+a_{3}z'_{3}+a_{5}z'_{5}&=&a_{2}z'_{2}+a_{4}z'_{4}+a_{6}z'_{6}\\
\lambda_{W}&=&a_{1}z''_{1}+a_{3}z''_{3}+a_{5}z''_{5}&=&a_{2}z''_{2}+a_{4}z''_{4}+a_{6}z''_{6}
\end{array}
\right.
\end{equation}
where $z_i$, $z_i'$ and $z_i''$ denote ordered spectra of $U_1$, $U_2$
and $W$, respectively. 
It is now clear that for a generic choice of $U_1$ and $U_2$
(which implies $W=U_1^{\dagger}U_2$), 
such a system has no solutions.
However,  if  both diagonal matrices are of the special form (\ref{daiagU}),
there exists a solution of the problem. The  weights $a_i$ satisfy 
\begin{equation}
\left\{
\begin{array}{ccccc}
a_{1}&=&a_{2}&=&1+\displaystyle\frac{1}{-1+\cos\alpha}\\
\\
a_{3}&=&a_{4}&=&\displaystyle\frac{1}{2-2\cos\alpha}\\
\\
a_{5}&=&a_{6}&=&\displaystyle\frac{1}{2-2\cos\alpha}
\end{array}\right. 
\label{eq:wspolczynniki}
\end{equation}
and imply the following compression values  
\begin{equation}
\left\{
\begin{array}{ccc}
\lambda_{U_{1}}&=&0\\
\lambda_{U_{2}}&=&0\\
\lambda_{W} &= & -1-2\cos\alpha
\end{array}\right. .
\label{eq:wspkompnum}
\end{equation}
Due to the symmetry of the problem the latter number $\lambda_W$ is real.

Substituting the weights (\ref{eq:wspolczynniki})
into (\ref {eq:wektfalowe})
we get an explicit form  (\ref{eq:p2}) of the subspace $C$.
It is now easy to check that this subspace satisfies 
simultaneously all three equations (\ref{eq:rkompresji})
with compression values given by (\ref{eq:wspkompnum}),
hence it provides a two dimensional error correction code
for this noise model.
This solution is correct for any unitaries $U_{1}$ and $U_{2}$ having any
set of eigenvectors  $|\varphi_{i}\rangle,\ i=1,\ldots,6$ and spectra  
given by (\ref{daiagU}) and parameterized by phases $\alpha$ and $\xi$.

The above construction can be generalized for 
a tri--unitary noise model acting on larger system
of size $N=3\times K$ \cite{Ma07}.
An error correction code of size $K$ exists in this case,
if matrices $U_1$ and $U_2$ 
have the tensor product form (\ref{eq:rtens}),
where
$U_A={\rm diag} \Big\{ 1,  e^{i\alpha},  e^{-i\alpha} \Big\} $
as before and
$U_B={\rm diag} \Big\{ 1,  e^{i\xi_2}, e^{i\xi_3},\dots ,e^{i\xi_{K}} \Big\} $.
The code subspace  $C=\sum_{i=1}^{K} |\psi_i\rangle \langle \psi_i|$ 
is then obtained 
in an analogous way, by representing the Hilbert space
as a direct product of $K$  subspaces of dimension three each
and constructing each state $|\psi_i\rangle$
as a coherent superposition of three eigenstates of $U_1$
corresponding to a triple of eigenvalues 
$z_l,z_{l+K}$ and $z_{l+2K}$ for $l=1, \dots K$. 
Note that the code space constructed here 
for the bipartite system does not have the tensor product structure, 
since it  is spanned  by entangled states  (\ref{eq:wektfalowe}).

\section{Conclusions}\
This paper concerns finite dimensional instruments or Kraus measurements,
acting on a quantum system with Hilbert space $\H$.
We have proved a version of Heisenberg's Principle,
which connects `darkness' to `protection' of a subspace $\L$ of $\H$.
`Darkness' expresses the lack of visibility of the information contained
in $\L$ from the measurement outcome,
and `protection' the degree to which this information remains present
in the quantum system.
Complete darkness corresponds to complete recoverability of information
as in error correction codes.

We have presented examples of darkness and protection:
instruments arising from random external fields,
arbitrary rank 2 channels, and biased permutation channels.
Bi-unitary noise models were analyzed recently in regard to their
error correction properties in \cite{CKZ06,CHKZ07}. 
Here we have also considered tri-unitary noise.
For a a certain class of 
tri-unitary noise models acting on a $3 \times K$ quantum system,
we have explicitly constructed an error correction code of size $K$. 
Although  this particular noise model might be considered as not very realistic, 
we tend to believe that the technique proposed can be 
applied to a broader class of quantum systems.

%



%

\section{Acknowledgements}

We enjoyed fruitful discussions with J. A. Holbrook, P. Horodecki and D. Kribs.
We acknowledge financial support by the Polish Research Network LFPPI
and by the European Research Project SCALA.



\begin{thebibliography}{99} 
 

\bibitem{DG97} L.-M. Duan and G.-C. Guo,
{\sl Phys. Rev. Lett.} {\bf 79}, 1953 (1997).

\bibitem{ZR97} P. Zanardi and M. Rasetti,
{\sl Phys. Rev. Lett.} {\bf 79}, 3306 (1997).

\bibitem{LCW98a} D.A. Lidar, I.L. Chuang, and K.B. Whaley,
{\sl Phys. Rev. Lett.} {\bf 81}, 2594 (1998).

\bibitem{SL05} A. Shabani and D. A. Lidar,
{\sl Phys. Rev.} {\bf  A 72}, 042303 (2005).

 \bibitem{BDSW96a}
 {C.~H.} Bennett, D.~P. DiVincenzo, J.~A. Smolin,
 and W.~K. Wootters
 {\sl Phys. Rev.} {\bf  A 54}, 3824 (1996).

\bibitem{KL97} E.~Knill and R.~Laflamme,
    {\sl Phys. Rev.} {\bf A 55}, 900 (1997).


\bibitem{MK06} H. Maassen and B. K{\"u}mmerer,
Purification of quantum trajectories.
In: Institute of Mathematical Statistics, Lecture Notes -- Monograph Series
Vol. 48 (eds. Dee Denteneer, Frank den Hollander, Evgeny Verbitsky), pp.
252-261  (2006) and also quant-ph/0505084

\bibitem{Kr71} K. Kraus, 
 General state changes in quantum theory,
  {\sl Ann. Phys. } {\bf 64}, 311 (1971).

\bibitem{CKZ06b} M.-D. Choi, D. W. Kribs, and  K. {\.Z}yczkowski,
Higher-Rank Numerical Ranges and Compression Problems, 
{\sl Lin. Alg. Appl.} {\bf  418}, 828-839 (2006)  

\bibitem{CKZ06} M.~D.~Choi, D.~W.~Kribs and K.~{\.Z}yczkowski,
  Quantum error correcting codes
   from the compression formalism,   
  {\sl Rep. Math. Phys.} {\bf 58}, 77 (2006).


\bibitem{Cho75a} M.-D. Choi,
Completely positive linear maps on complex matrices,
{\sl Linear Alg. Appl.} {\bf 10}, 285 (1975).

\bibitem{BZ06}   I. Bengtsson  and K. {\.Z}yczkowski,
{\sl Geometry of Quantum States: An Introduction to Quantum
Entanglement}, Cambridge University Press, Cambridge 2006.

\bibitem{KNMR} C. King, K. Matsumoto, M. Nathanson, M.~B.~Ruskai,
Properties of Conjugate Channels with Applications to
Additivity and Multiplicativity,
{\sl  Markov Process Related Fields} {\bf  13}, 391-423 (2007).


\bibitem{DaL} E.~B. Davies, J.~T.~Lewis,
An Operational Approach to Quantum Probability,
{\sl Comm. Math. Phys.} {\bf 17}, 239-260 (1970).


\bibitem{Hei27} W. Heisenberg,
\"Uber den anschaulichen Inhalt der Quantentheoretischen Kinematik
und Mechanik,
{\sl Z. Phys. \bf 43}, 172-198 (1927).


\bibitem{Rob29} H. Robertson,
The uncertainty principle,
{\sl Phys. Rev. \bf 34}, 163-164 (1929).

\bibitem{Wer04} R.F. Werner,
The uncertainty relation for joint measurement of position and momentum,
{\sl Quant. Inf. Comp. \bf4}, 546-562 (2004).

\bibitem{Jan06}B. Janssens,
Unifying decoherence and the Heisenberg principle,
{\tt www.arxiv.org/quant-ph/0606093}.

\bibitem{Gi89} N. Gisin,
Stochastic quantum dynamics and relativity,
{\sl Helv. Phys. Acta}  {\bf 62}, 363  (1989).

\bibitem{HJW93} L. P. Hughston and  R. Jozsa and W.~K.~Wootters,
A complete classification of quantum ensembles 
having a given density matrix,
{\sl Phys. Lett.} {\bf A~183}, 14  (1993).

\bibitem{LBW99} D.A. Lidar, D. Bacon, and  K.B. Whaley,
{\sl Phys. Rev. Lett.} {\bf 82}, 4556 (1999).

\bibitem{KPZ08}  D.W. Kribs, A. Pasieka, K. {\.Z}yczkowski, 
Entropy of a quantum error correction code,
{\sl Open Syst. Inf. Dyn.} {\bf 15}, 329-343 (2008)

\bibitem{Bh97} R. Bhatia, {\sl Matrix Analysis},
 Springer Verlag, New York 1997.

\bibitem{CHKZ07} M.~D.~Choi,  J.~A.~ Holbrook, D.~W.~Kribs
 and K.~{\.Z}yczkowski,
{\sl Operators and Matrices} {\bf 1}, 409 (2007).

\bibitem{LS08} C.-K. Li, and N.-S. Sze,
 Canonical forms, higher rank numerical ranges, totally isotropic subspaces, and matrix equations,
{\sl Proc. Amer. Math. Soc.} {\bf 136}, 3013-3023 (2008).

\bibitem{Wo08}  H. Woerdeman,
 The higher rank numerical range is convex,
 {\sl  Lin. \& Multilin. Algebra} {\bf 56},  65-67 (2008).

\bibitem{CGHK08}  M.-D. Choi, M. Giesinger J.A. Holbrook, D.W. Kribs,
 Geometry of higher-rank numerical ranges,
{\sl Lin. \& Multilin. Algebra} {\bf 56},  53-64 (2008).

\bibitem{LPS07} C.-K. Li, Y.-T. Poon and N.-S. Sze,
 Condition for the higher rank numerical range to be non-empty,
{\sl Lin. \& Multilin. Algebra} {\bf 57}, 365-368 (2009).

\bibitem{LPS07b} C.-K. Li, Y.-T. Poon and N.-S. Sze,
  Higher rank numerical ranges and low rank perturbations of quantum channels, 
 {\sl  J. Math. Analysis  Appl.} {\bf  348}, 843-855  (2008)


\bibitem{AL87}  R.~Alicki and K.~Lendi, {\sl  Quantum Dynamical Semigroups and
Their Applications},  LNP 286, Springer, Berlin (1987)

\bibitem{Ma07} K. Majgier,
Quantum error correction codes   
for unitary models of noise ({\sl in Polish}),
Master thesis, Jagiellonian University, Cracow, June 2007;
see http://chaos.if.uj.edu.pl/$\sim$karol/prace/Majgier07.pdf


\end{thebibliography}
\end{document}